\documentclass[epj]{svjour}
\usepackage{latexsym}
\usepackage{graphicx}
\usepackage{color}
\usepackage{amssymb}
\usepackage{amsmath}
\usepackage{hyperref}
\usepackage{float}
\usepackage{url}
\usepackage{subfig}
\begin{document}
\title{Nuclear Modification Factor Using Tsallis Non-extensive Statistics}
\author{Sushanta Tripathy\inst{1}, Trambak Bhattacharyya\inst{2}, Prakhar Garg\inst{1,}\thanks{\emph Present Address: Department of Physics and Astronomy, Stony Brook University, SUNY, Stony Brook, NY 11794-3800, USA}, Prateek Kumar\inst{1}, Raghunath Sahoo\inst{1,}\thanks{\emph e-mail: Raghunath.Sahoo@cern.ch (corresponding author)},   \and Jean Cleymans\inst{2}
}                     
%
%
\institute{Discipline of Physics, School of Basic Sciences, Indian Institute of Technology Indore, Simrol, M.P.- 453552, India 
\and UCT-CERN Research Centre and Department of Physics,
University of Cape Town, Rondebosch 7701, South Africa}
\date{\today}
%
\abstract{ The nuclear modification factor is derived using
Tsallis non-extensive statistics in relaxation time approximation. 
The variation of the nuclear modification factor with transverse momentum for 
different values of the non-extensive parameter, $q$, is also observed. The experimental data from RHIC and LHC are 
analysed in the framework of Tsallis
non-extensive statistics in a relaxation time approximation. 
It is shown that the proposed approach explains the $R_{AA}$ of all particles over
a wide range of transverse momentum but does not seem to describe the rise in $R_{AA}$ at very high
transverse momenta.} 
\authorrunning{Sushanta Tripathy {\it et al.}}
\titlerunning{Nuclear Modification Factor...}
\maketitle
%

\section{Introduction}
One of the major goals of studying heavy-ion collisions at high-energies is to
search for a deconfined state of quarks and gluons, also known as quark gluon
plasma (QGP), and to study its properties. The
bulk properties of the QGP are governed by light quarks and gluons. The heavy quarks act as probes for QGP properties due to the fact that they
witness the entire plasma evolution as they are produced in the initial hard
scattering and endure until hadronization. Also, as their time scale of
thermalization is longer than that of light quarks, they can retain the entire
interaction history more effectively. Similarly, the energy loss of light
quarks becomes important to study the flavor dependence of the energy loss in heavy-ion collisions.

The energy loss is more for high-$p_{T}$ heavy and light quark flavors
\cite{Arsene:2006ts,Bjorken:1982qr,Pluemer:1995dp,Baier:1994bd} due to
interaction with the medium. Finally they appear as constituents of
hadrons. The propagation of energetic quarks through the medium has been treated
as Brownian motion which is described by means of Fokker-Planck equation. In
this equation, the interaction is encoded in drag and diffusion coefficients.
Many theoretical efforts have been made using Fokker-Planck equation to
reproduce the experimentally observed value of $R_{AA}$ of heavy and light
quarks \cite{Alam:2006bu,Svetitsky:1987gq,GolamMustafa:1997id,vanHees:2005wb,vanHees:2007me,Das:2009vy,Das:2015ana,albericoraa,Mazumder:2011nj}.
Also, there have been studies to elucidate the dominant mode of energy loss
\cite{Alam:2006bu} or the flavor dependence of energy loss
\cite{Zakharov:2012fp,Kopeliovich:2010aia}. But, till date the issues are far
from being settled \cite{Kolbe:2015rvk}.

The nuclear modification factor ($R_{AA}$) is a measure of the modification of particle production.
It can be represented as
\begin{eqnarray}
\label{eq1}
R_{AA}=\frac{f_{fin}}{f_{in}},
\label{raatheo}
\end{eqnarray}
where $f_{in}$ is the distribution of the highly energetic particles immediately
after their formation and $f_{fin}$ is the distribution of the particles
after the interaction with the medium.  

$R_{AA}$ is defined as
\begin{eqnarray}
 \label{eq2}
 R_{AA}(p_{T})=\frac{(1/N_{AA}^{evt})d^{2}N_{AA}/dydp_{T}}{(\langle
N_{coll}\rangle/\sigma_{NN}^{inel})\times d^2\sigma_{pp}/dydp_T},
 \label{raaexpt}
\end{eqnarray}
where $d^{2}N_{AA}/dydp_{T}$ is the yield in A+A collisions, $\langle N_{coll}\rangle$ is the
number of binary nucleon-nucleon collisions averaged over the impact parameter range of the 
corresponding centrality bin calculated by Glauber Monte-Carlo simulation
\cite{Glauber:1970jm}. $\sigma_{NN}^{inel}$
is the inelastic cross section and $d^2\sigma_{pp}/dydp_T$ is the differential
cross section for inelastic $p+p$ collisions. $N_{AA}^{evt}$ is the number of events in A+A collisions. 
If, $R_{AA} = 1$, this indicates
that A+A collisions are mere superposition of scaled $p+p$ collisions. A deviation
of $R_{AA}$ from unity indicates the medium modification. It has been observed
that high-$p_T$ particle yields in Au+Au and Pb+Pb collisions at RHIC and LHC
are suppressed as compared to $p+p$ collisions \cite{Adler:2003qi,Aamodt:2010jd},
which suggests the formation of a dense medium.

In this work, we represent the initial distribution of the energetic particles
with the help of Tsallis power law distribution parameterized by
the Tsallis $q$ parameter and the Tsallis temperature $T$, remembering the fact
that their genesis is due to very hard scatterings. We plug the initial
distribution ($f_{in}$) in Boltzmann Transport Equation (BTE) and solve it with
the help of Relaxation Time Approximation (RTA) of the collision term to find
out the final distribution ($f_{fin}$). Hence, the ratio in Eq. \ref{raatheo} expressible in terms of
$q$, $T$ and relaxation time $\tau$ can be computed and compared with the
experimentally observed values.

The paper is organized as follows. In section \ref{raa}, the nuclear modification
factor is derived using RTA of the BTE and the expression for nuclear
modification factor is derived. In section \ref{results}, fits to the
experimental data using the proposed model along with results and
discussions are presented; and lastly, we summarize our findings in
section \ref{summary}.

\section{Nuclear Modification Factor in Relaxation time approximation (RTA)}
\label{raa}

The evolution of the particle distribution owing to its interaction
with the medium particles can be studied through Boltzmann transport
equation,
\begin{eqnarray}
\label{eq3}
 \frac{df(x,p,t)}{dt}=\frac{\partial f}{\partial t}+\vec{v}.\nabla_x
f+\vec{F}.\nabla_p
f=C[f],
\end{eqnarray}
where $f(x,p,t)$ is the distribution of particles which depends on position, momentum
and time. $\bf{v}$ is the velocity and $\bf{F}$ is the external force. $\nabla_x$ and $\nabla_p$ are the partial derivatives
with respect to position and momentum, respectively. $C[f]$ is the collision
term which encodes the interaction of the probe particles with the medium. The Boltzmann Transport Equation has earlier also been used in relaxation time approximation to study the time evolution of temperature fluctuation in a non-equilibrated system \cite{trambak-rns}.

Assuming homogeneity of the system ($\nabla_x f=0$) and absence of external
force ($F=0$), the second and third terms of the above equation become zero and Eq. \ref{eq3} becomes,
\begin{eqnarray}
 \label{eq4}
  \frac{df(x,p,t)}{dt}=\frac{\partial f}{\partial t}=C[f]
\end{eqnarray}


In relaxation time approximation \cite{balescu,Florkowski:2016qig}, the collision term can be expressed as,
\begin{eqnarray}
\label{eq5}
 C[f] =-\frac{f-f_{eq}}{\tau}
 \label{colltermrta}
\end{eqnarray}
where $f_{eq}$ is Boltzmann local equilibrium
distribution characterized by a temperature $T_{eq}$.
$\tau$ is the relaxation time, the time taken by a non-equilibrium
system to reach equilibrium. With the ansatz in Eq. \ref{colltermrta}, Eq.
\ref{eq4} becomes,
\begin{eqnarray}
 \label{eq6}
  \frac{\partial f}{\partial t}=-\frac{f-f_{eq}}{\tau}
\end{eqnarray}
Solving the above equation in view of initial conditions i.e. at $t=0, f=f_{in}$
and at $t=t_f, f=f_{fin}$; leads to,
\begin{eqnarray}
 \label{eq7}
 f_{fin}=f_{eq}+(f_{in}-f_{eq})e^{-\frac{t_f}{\tau}},
\end{eqnarray}
where $t_f$ is the freeze-out time. Using Eq.\ref{eq7}, the nuclear modification factor can be expressed as,
\begin{eqnarray}
\label{eq8}
R_{AA}=\frac{f_{fin}}{f_{in}}=\frac{f_{eq}}{f_{in}}+\left ( 1-\frac{f_{eq}}{f_{in}}\right )e^\frac{-t_{f}}{\tau}
\end{eqnarray}
Eq. \ref{eq8} is the derived nuclear modification factor after incorporating
relaxation time approximation, which is the basis of our analysis in the present paper. It involves the (power law-like) initial
distribution and the equilibrium distribution. In this analysis, the initial
distribution is parameterized using the thermodynamically consistent Tsallis
distribution \cite{worku}.


Tsallis statistics is widely used to analyse the detected particle spectra
in high-energy collisions starting from $e^{+}e^{-}$, $p+p$ to heavy-ions
\cite{Thakur:2016boy,e+e-,R1,R2,R3,ijmpa,plbwilk,marques,STAR,PHENIX1,PHENIX2,ALICE_charged,ALICE_piplus,CMS1,CMS2,ATLAS,ALICE_PbPb,Bhattacharyya:2015hya}. Also,
as the system stays away from thermal equilibrium during
the formation of highly energetic particles immediately after the collision,
the particle distribution can be parameterized with the help of Tsallis
distribution. A thermodynamically consistent non-extensive Tsallis distribution
function, to be used as the initial distribution, is given by, \cite{worku}
\begin{eqnarray}
\label{eq9}
f_{in}=\frac{gV}{(2\pi)^2} p_T m_T
\left[1+{(q-1)}{\frac{m_T}{T}}\right]^{-\frac{q}{q-1}},
\end{eqnarray}
which is used for studying the particle distribution stemming from the proton-proton collisions
as discussed in Ref. \cite{worku}.

The Boltzmann equilibrium distribution is given by

\begin{equation}
 f_{eq}=\frac{gV}{(2\pi)^2} p_T m_T e^{-\frac{m_T}{T_{eq}}}
 \label{boltz}
\end{equation}

Here, $V$ is the system volume, $m_{\rm T}=\sqrt{p_T^2+m^2}$ is the transverse
mass and $q$ is the non-extensive parameter,
which measures the degree of deviation from equilibrium. Using Eqs. \ref{eq9}
and \ref{boltz} (both for mid-rapidity and for zero chemical potential)
nuclear modification factor can be expressed as,

\begin{eqnarray}
 \label{eq10}
R_{AA}= \frac{e^{-\frac{m_T}{T_{eq}}}}{(1+(q-1)\frac{m_T}{T})^{-\frac{q}{q-1}}}+\nonumber\\\left[1-\frac{e^{-\frac{m_T}{T_{eq}}}}{(1+(q-1)\frac{m_T}{T})^{-\frac{q}{q-1}}}\right]e^{-\frac{t_f}{\tau}}
\end{eqnarray}

This will be compared to experimental results in the next section.

\section{Results and Discussion}
\label{results}
\begin{figure}[ht!]
\begin{center}
\resizebox{0.45\textwidth}{!}{ \includegraphics{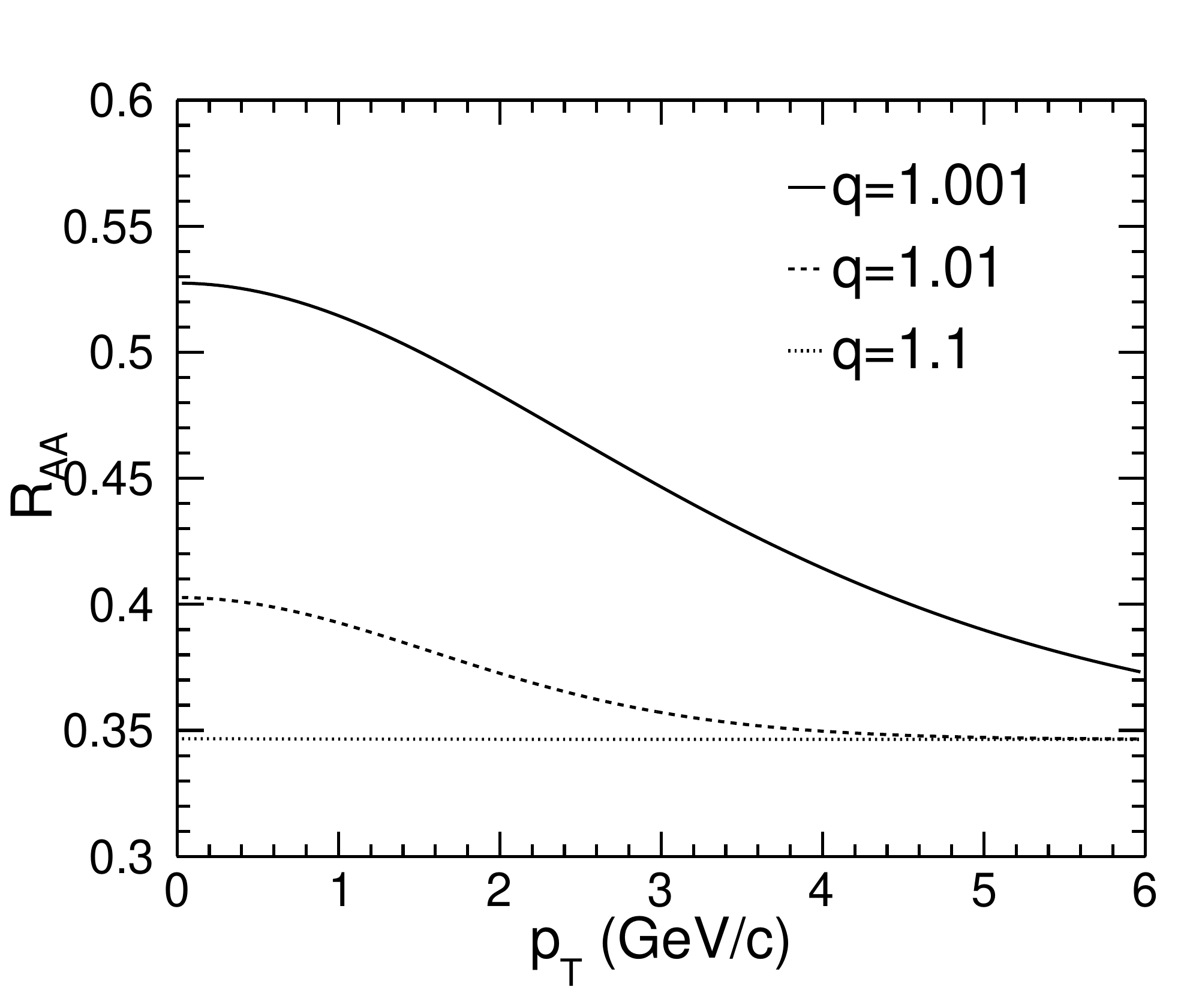}}
\caption{Nuclear modification factor versus $p_T$ for different values of the
non-extensive parameter using Tsallis Boltzmann distribution as shown in
Eq.\ref{eq10}. Here $m = 3.096$ GeV, $T_{eq}=0.16$ GeV, $T=0.17$ GeV and
$t_f/\tau = 1.06$.}
\label{fig1}
\end{center}
\end{figure}

To illustrate the formula given in Eq.~\ref{eq10} we take as an example the case of
the $J/\psi$ particle. The variation of nuclear modification factor with transverse momentum for
different values of non-extensive parameter is plotted in Fig. \ref{fig1}. For
this figure,
$m = 3.096$ GeV, $T_{eq}=0.16$ GeV, $T=0.17$ GeV and $t_f/\tau = 1.06$ are taken.
It is observed that for higher
values of $q$, $R_{AA}$ decreases for all the values of $p_T$. This suggests
that when the initial distribution remains closer to equilibrium, the suppression becomes
less. The model presented here fails to describe the increase in $R_{AA}$ at larger
transverse momenta, which is seen in experimental data for light flavor hadrons.

We now proceed to the more detailed analysis of the experimental data with the
model proposed above. Keeping all
the parameters free, we fit the spectra for different particles in different  centralities for Pb+Pb and Au+Au collisions using TMinuit
class available in ROOT library \cite{root} to get a convergent solution. The
convergent solution is obtained by $\chi ^2$ minimization technique. Here $T$,
$q$ and ${t_f}/{\tau}$ are the fitting parameters for the experimental data. The
equilibrium temperature $T_{eq}$ is fixed to 160 MeV throughout the analysis.

In our analysis, it is observed that the fitting of low $p_T$ for light flavor
particles fails due to the reason that it involves different physical processes such
as regeneration, coalescence, shadowing etc., which are out for the scope of present
formalism. But in heavy flavor particles, no such processes are involved at the
discussed energies. Thus our proposed model explains successfully the heavy
flavor $R_{AA}$ data. Also, for very high-$p_T$, our model fails to
explain the increase in $R_{AA}$ (most prominent for $K^{\pm}$ in
Fig. \ref{fig3}).

\begin{figure}[ht!]
\begin{center}
\resizebox{0.40\textwidth}{!}{ \includegraphics{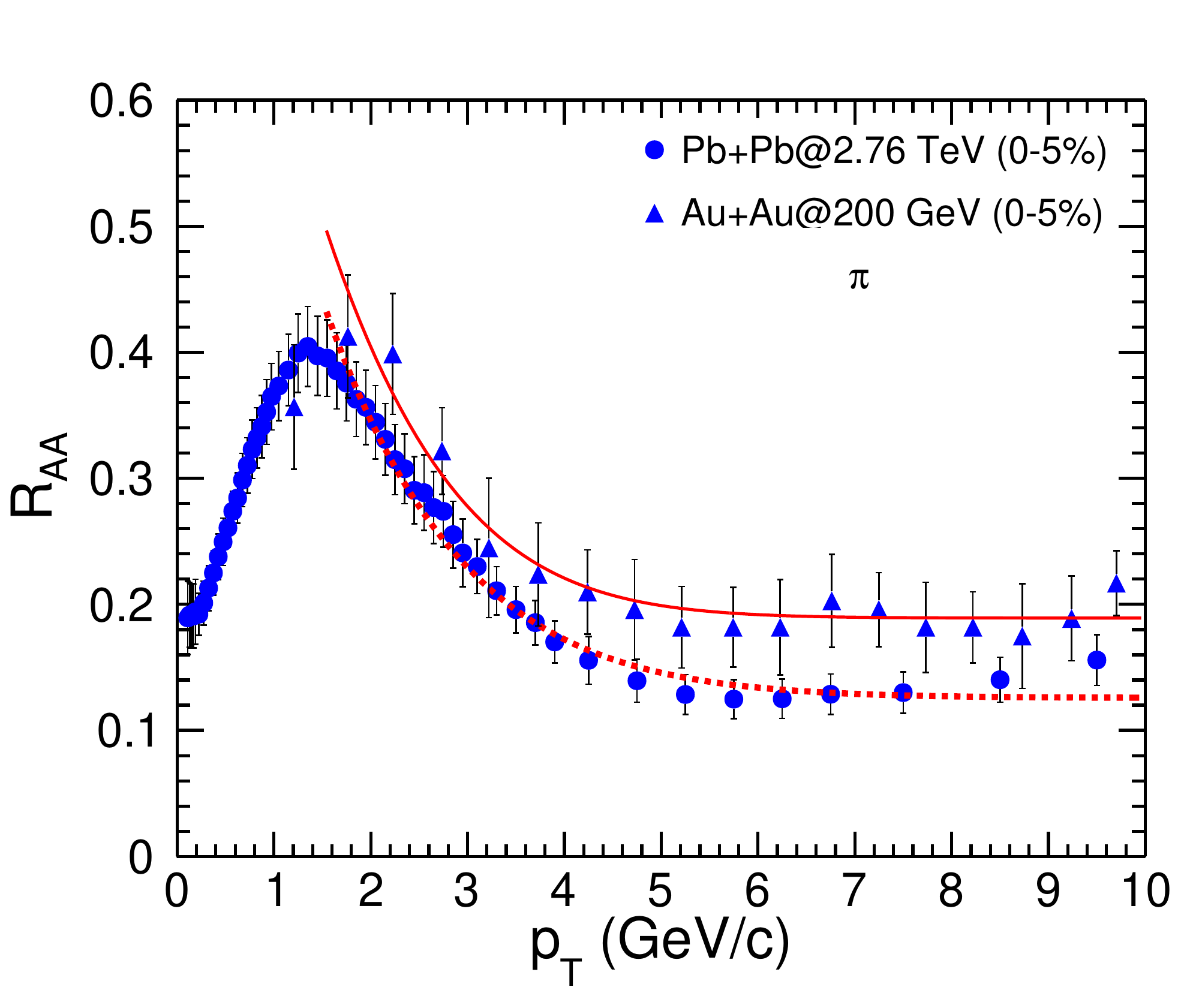}}
\caption{(Color Online) Fitting of experimental data for nuclear modification
factor with our proposed model (Eq.\ref{eq10}) for
$\pi^0 $\cite{Adare:2008qa} (blue triangles) in Au+Au collisions at
$\sqrt{\mathrm{s}_{NN}}$= 200 GeV and $\pi^ {+}+ \pi^{-} $\cite{Abelev:2014laa}
in Pb+Pb collisions at 
$\sqrt{\mathrm{s}_{NN}}$= 2.76 TeV (blue dots). The solid red line shows the
fitting for blue triangles and the dotted red line shows the fitting for blue
dots.}
\label{fig2}
\end{center}
\end{figure}

\begin{figure}[ht!]
\begin{center}
\resizebox{0.40\textwidth}{!}{ \includegraphics{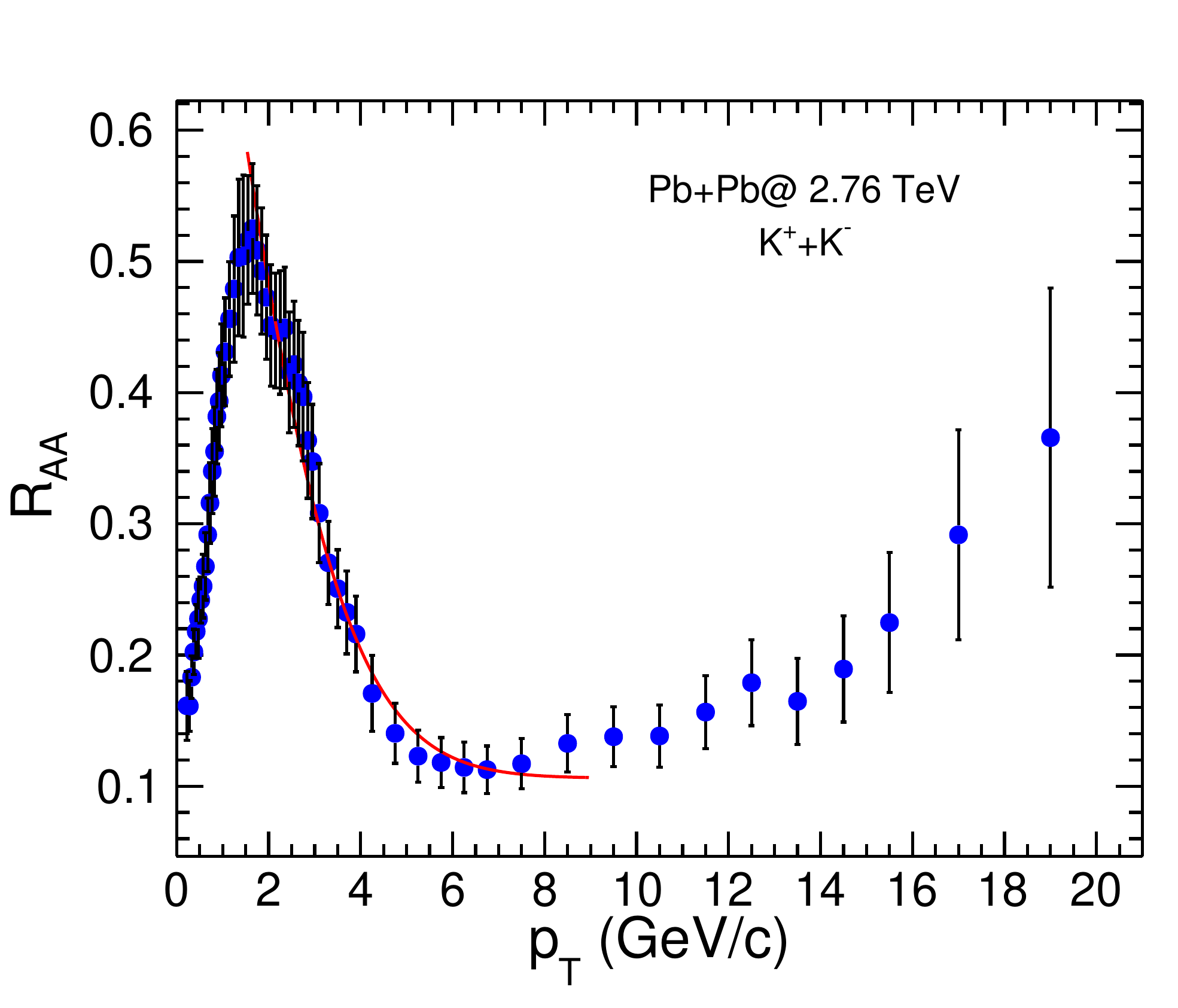}}
\caption{(Color Online) Fitting of experimental data for nuclear modification
factor with our proposed model(Eq.\ref{eq10}) for
$K^{+}+K^{-} $\cite{Abelev:2014laa} in
most central Pb+Pb collisions at 
$\sqrt{\mathrm{s}_{NN}}$= 2.76 TeV (blue dots). The solid red line shows the
fitting for blue dots.}
\label{fig3}
\end{center}
\end{figure}

Fig. \ref{fig2} shows
the fitting of experimental data using Eq.\ref{eq10} for $\pi^0$ meson in
most central Au+Au collisions
at $\sqrt{\mathrm{s}_{NN}}$= 200 GeV. The derived expression for $R_{AA}$ fits the data in
intermediate to
high-$p_T$ range. Also, in Fig. \ref{fig2} we show the fitting of experimental
data for $\pi^ {+}+ \pi^{-}$ in most central Pb+Pb
collisions at $\sqrt{\mathrm{s}_{NN}}$= 2.76 TeV. The model considered here, fits
the data from intermediate
to high-$p_T$ range. The fitting parameters are shown in table \ref{table1}
along with the $\chi^2/ndf$ values.

Similarly, in Fig. \ref{fig3} we fit the experimental data for $K^{+}+K^{-}$ in most central Pb+Pb collisions
at $\sqrt{\mathrm{s}_{NN}}$= 2.76 TeV. The proposed model fits the data for
intermediate $p_T$ range. The derived expression of $R_{AA}$ could not fit the
data in high-$p_T$ range as an enhancement is observed in high-$p_T$.  
The fitting parameters are shown table \ref{table1} along with the
$\chi^2/ndf$ values.

\begin{figure}[ht!]
\begin{center}
\resizebox{0.40\textwidth}{!}{ \includegraphics{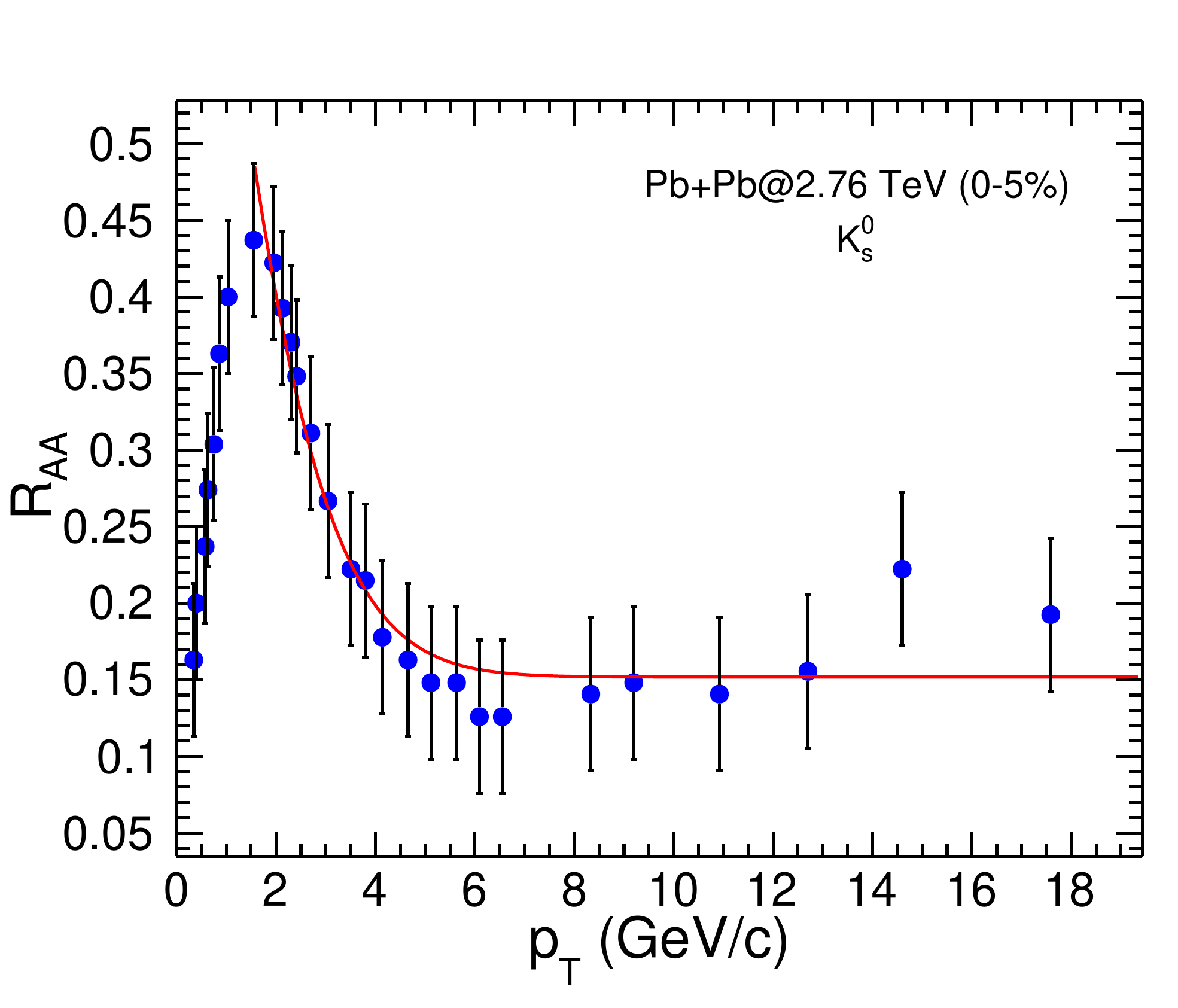}}
\caption{(Color Online) Fitting of experimental data for nuclear modification
factor with our proposed model (Eq.\ref{eq10}) for $K_{S}^{0}$ in most
central Pb+Pb collisions at 
$\sqrt{\mathrm{s}_{NN}}$= 2.76 TeV(blue dots). The solid red line shows the
fitting for blue dots.}
\label{fig4}
\end{center}
\end{figure}

Fig. \ref{fig4} shows
the fitting of experimental data for $K_S^0$ in most central Pb+Pb
collisions
at $\sqrt{\mathrm{s}_{NN}}$= 2.76 TeV. The proposed model fits the data for
intermediate to high-$p_T$ range. 
The fitting parameters are shown table \ref{table1} along with the
$\chi^2/ndf$ values.

\begin{figure}[ht!]
\begin{center}
\resizebox{0.40\textwidth}{!}{ \includegraphics{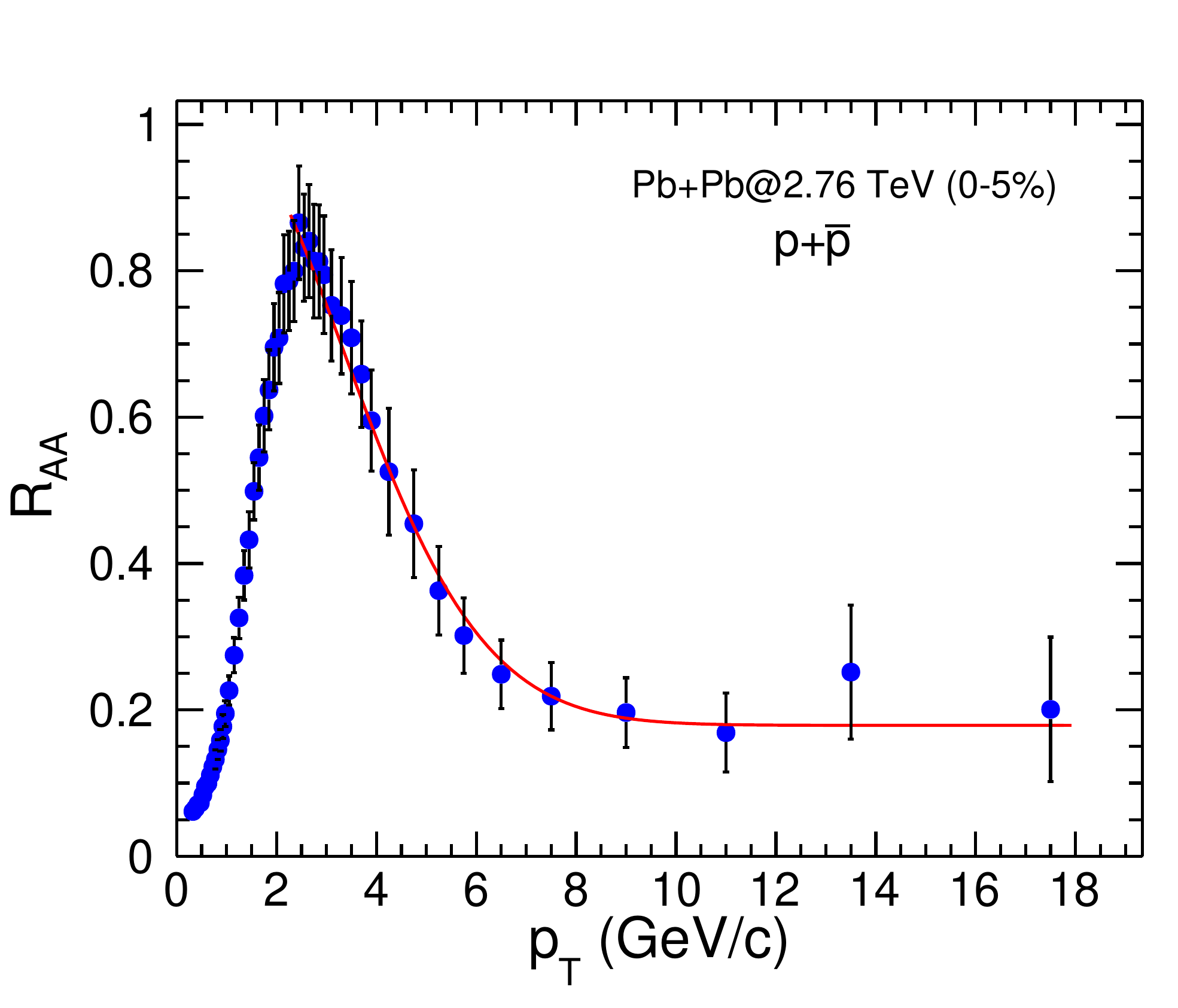}}
\caption{(Color Online) Fitting of experimental data for nuclear modification factor with our proposed model (Eq.\ref{eq10}) for $p+\bar p$ \cite{Abelev:2014laa} in 
most central Pb+Pb collisions at 
$\sqrt{\mathrm{s}_{NN}}$= 2.76 TeV (blue dots). The solid red line shows the fitting for blue dots.}
\label{fig5}
\end{center}
\end{figure}

Fig.\ref{fig5} shows
the fitting of experimental data for $p+\bar p$ in most central Pb+Pb collisions
at $\sqrt{\mathrm{s}_{NN}}$= 2.76 TeV. The proposed model fits the data for intermediate to high-$p_T$ range.
$\chi^2/ndf$ value and the fitting parameters are shown table \ref{table1}.

\begin{figure}[ht!]
\begin{center}
\resizebox{0.40\textwidth}{!}{ \includegraphics{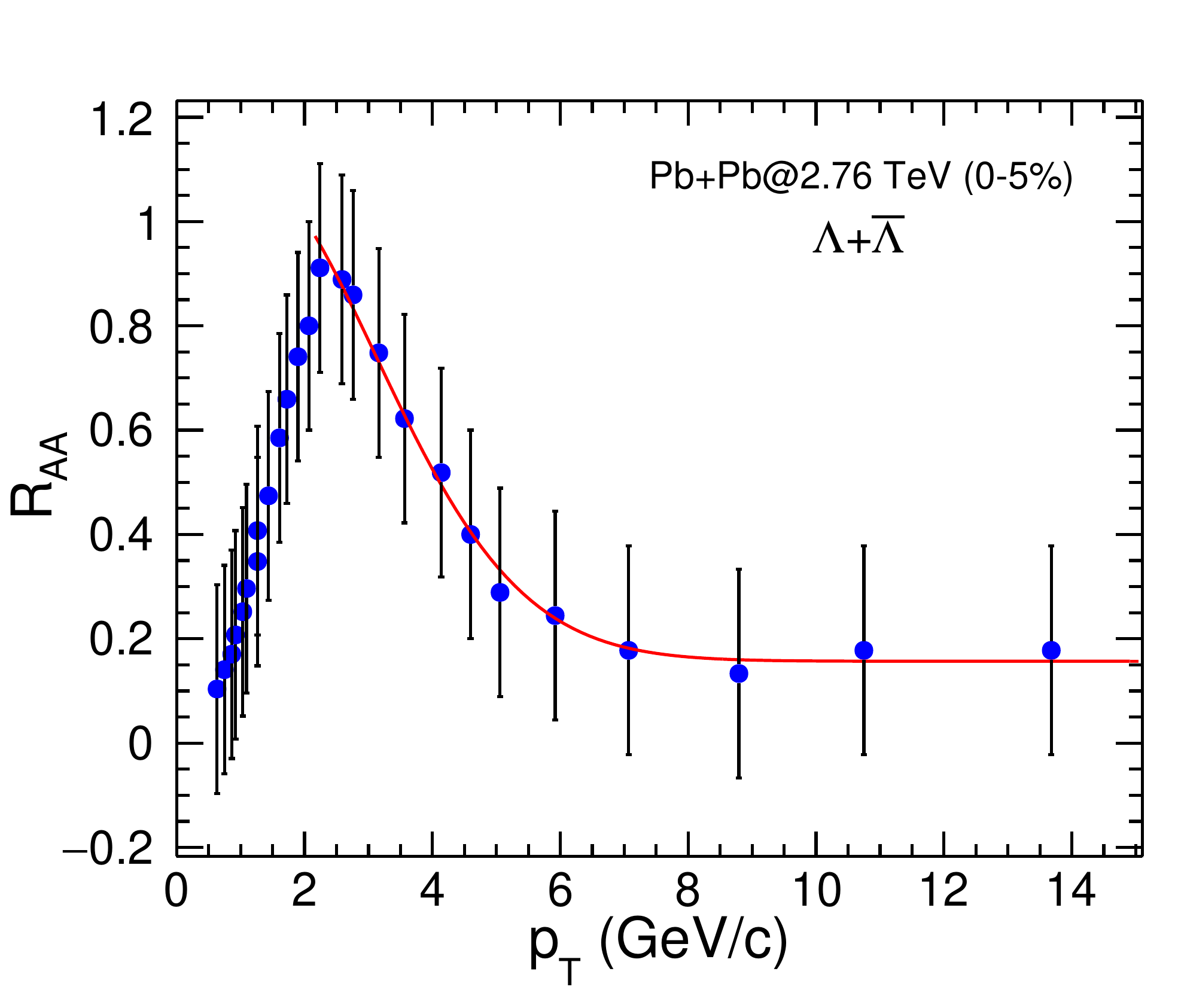}}
\caption{(Color Online) Fitting of experimental data for nuclear modification
factor with our proposed model (Eq.\ref{eq10}) for $\Lambda+\bar \Lambda$ in most central Pb+Pb collisions at 
$\sqrt{\mathrm{s}_{NN}}$= 2.76 TeV (blue dots). The solid red line shows the
fitting for blue dots.}
\label{fig6}
\end{center}
\end{figure}

Fig. \ref{fig6} shows
the fitting of experimental data using Eq.\ref{eq10} for $\Lambda+\bar
\Lambda$ in most central Pb+Pb collisions
at $\sqrt{\mathrm{s}_{NN}}$= 2.76 TeV. The proposed model fits the data from
intermediate to high-$p_T$ range starting from 2 GeV to 14 GeV.
$\chi^2/ndf$ value and the fitting parameters are shown table \ref{table1}.

\begin{figure}[ht!]
\begin{center}
\resizebox{0.40\textwidth}{!}{ \includegraphics{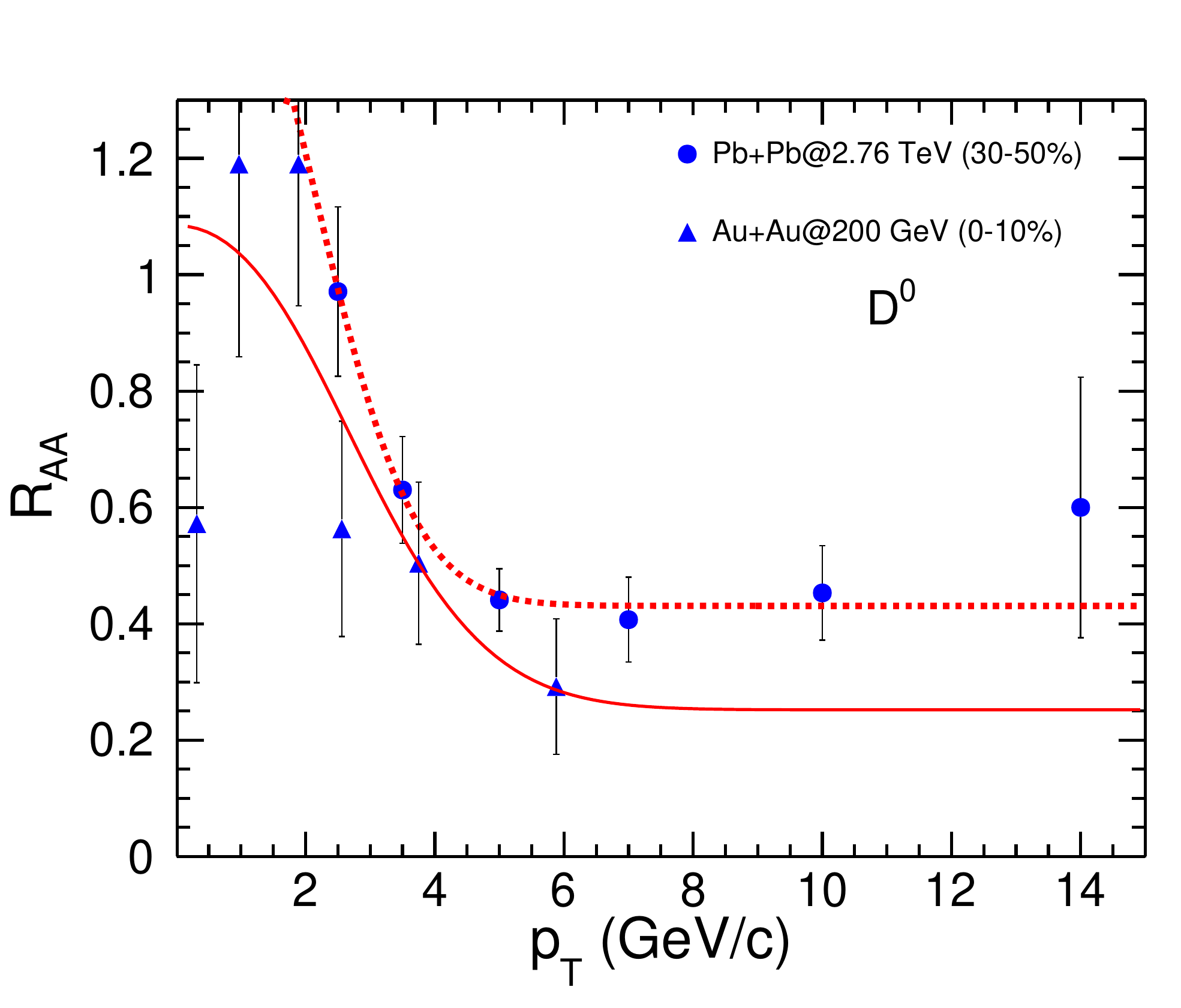}}
\caption{(Color Online) Fitting of experimental data for nuclear modification
factor with our proposed model (Eq.\ref{eq10}) for $D^0$ meson
\cite{Adamczyk:2014uip} (blue triangles)
in most central Au+Au collisions at $\sqrt{\mathrm{s}_{NN}}$= 200 GeV and $D^0$
meson \cite{Abelev:2014ipa} in (30-50)\% central Pb+Pb collisions at 
$\sqrt{\mathrm{s}_{NN}}$= 2.76 TeV (blue dots). The solid red line shows the
fitting for blue triangles and the dotted red line shows the fitting for blue
dots.}
\label{fig7}
\end{center}
\end{figure}

In Fig. \ref{fig7} we fit the experimental data for $D^0$ meson in most
central, (0-10)\% Au+Au collisions
at $\sqrt{\mathrm{s}_{NN}}$= 200 GeV. As less data points are available, the fitting parameters cannot
be established very well. Thus the ${\chi ^2}/{ndf}$ value is very high compared to other
particles, which can be seen table \ref{table1}.
Also in Fig. \ref{fig7} we show the fitting of experimental data for $D^0$ meson
for (30-50)\% central Pb+Pb collisions at $\sqrt{\mathrm{s}_{NN}}$= 2.76 TeV.
The proposed model fits the data accurately for all the $p_T$ ranges as
the enhancement is not involved which originate from regeneration through coalescence mechanism.
$\chi^2/ndf$ value and the fitting parameters are shown table \ref{table1}.

\begin{figure}[H]
\begin{center}
\resizebox{0.40\textwidth}{!}{ \includegraphics{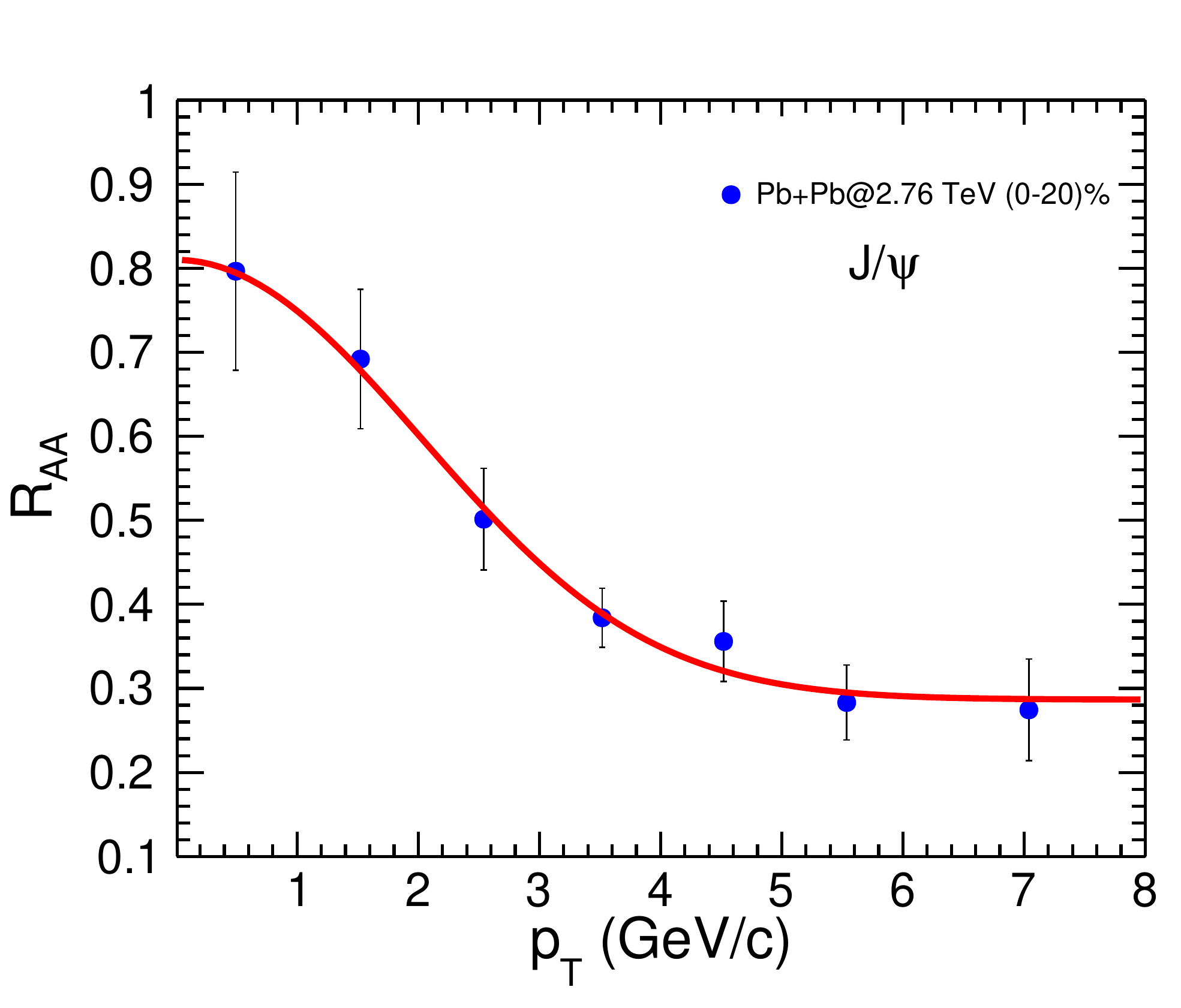}}
\caption{(Color Online) Fitting of experimental data for nuclear modification
factor with our proposed model (Eq.\ref{eq10}) for 
J/$\psi$ \cite{Abelev:2013ila} in Pb+Pb collisions at 
$\sqrt{\mathrm{s}_{NN}}$= 2.76 TeV (blue dots) with centrality (0-90)\%. The solid red line shows the
fitting to experimental data.}
\label{fig8}
\end{center}
\end{figure}

 In Fig. \ref{fig8} we show the fitting of experimental data for $J/\psi$ in
minimum bias Pb+Pb collisions (0-90\% centrality) at $\sqrt{\mathrm{s}_{NN}}$= 2.76 TeV.
The proposed model fits the data accurately for all $p_T$ ranges as the enhancement in $R_{AA}$ is not observed in experimental $J/\psi$ data. The $\chi^2/ndf$ value and the fitting parameters are shown table \ref{table1}. We have explicitly checked that the proposed model also explains the $J/\psi$ $R_{AA}$ for central Pb+Pb collisions at $\sqrt{\mathrm{s}_{NN}}$= 5.02 TeV. 


\begin{figure}[H]
\begin{center}
\resizebox{0.40\textwidth}{!}{ \includegraphics{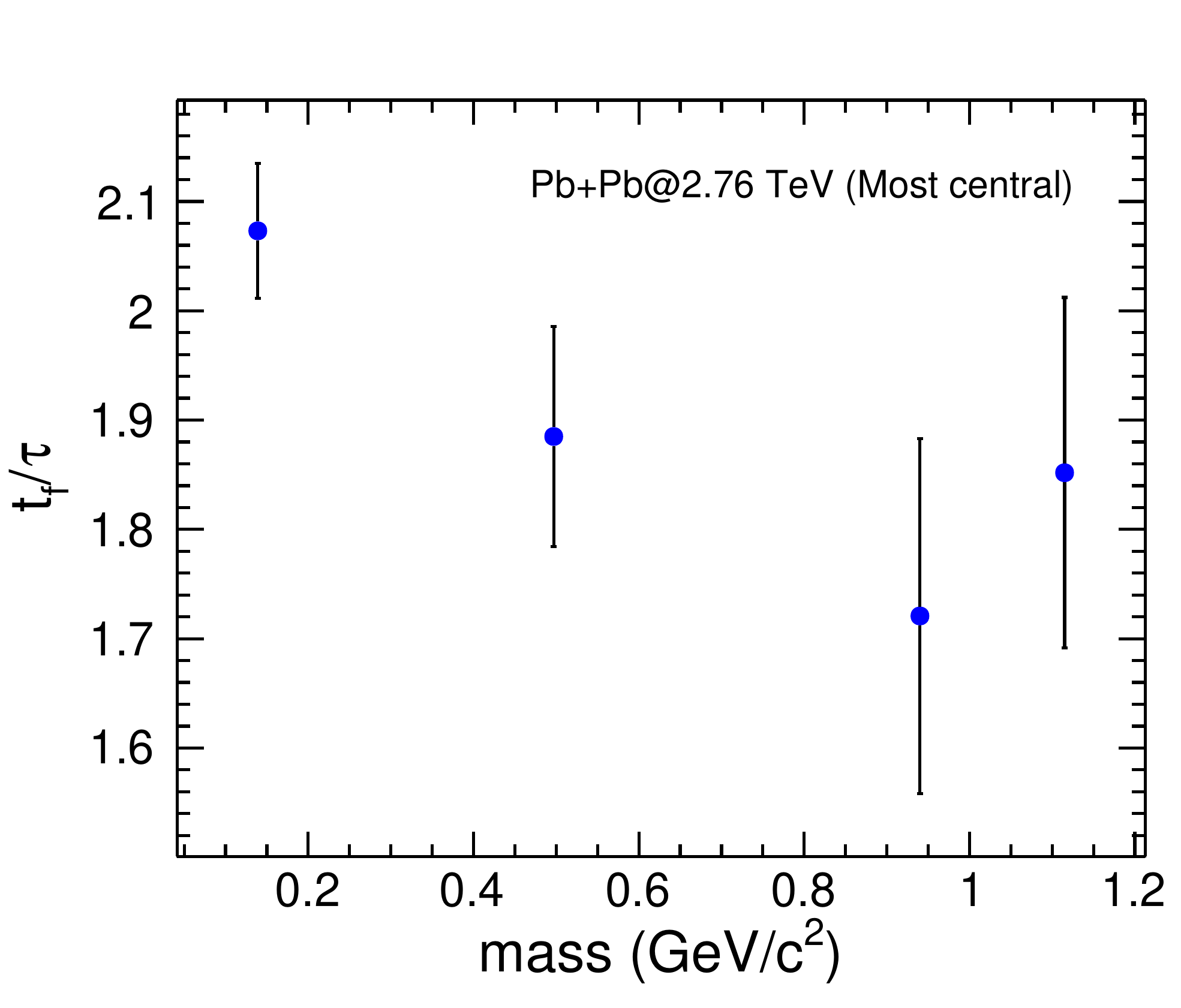}}
\caption{(Color Online) ${t_f}/{\tau}$ as a function of particle mass for most central Pb+Pb collisions at $\sqrt{\mathrm{s}_{NN}}$= 2.76 TeV. }
\label{fig9}
\end{center}

\end{figure}

Finally, we have shown the variation of ${t_f}/{\tau}$ with mass in Fig.
\ref{fig9}. It is observed that ${t_f}/{\tau}$ decreases with increasing the
particle mass, which suggests that
the heavy particles have more relaxation time compared to lighter particles.
Note here, that $t_f$ is the freeze-out time, and $\tau$ is the relaxation time, which differs from particle to particle, as the equilibration depends on the particle species and their interaction with the rest of the medium. Intuitively, the heavier the particle is, the more is the relaxation time. And hence, $t_f/\tau$ becomes less and the degrees of freedom are shared with other parameters used in the fit.

\section{Summary and Conclusion}
\label{summary}
In this work, we represent the initial distribution of the energetic particles
with the help of Tsallis power law distribution parameterized by
the Tsallis $q$ parameter and the Tsallis temperature $T$, remembering the fact
that their genesis is due to very hard scatterings. We plug the initial
distribution ($f_{in}$) in Boltzmann Transport Equation (BTE) and solve it with
the help of Relaxation Time Approximation (RTA) of the collision term to find
out the final distribution ($f_{fin}$). Hence, the ratio in Eq. \ref{raatheo} expressible in terms of
$q$, $T$ and relaxation time, $\tau$ can be computed and compared with the
experimentally observed values.
The variation of nuclear modification factor
with transverse momentum for different values of 
non-extensive parameter, is also observed. The suppression is found to be higher for a system with higher degree of deviation from equilibrium. 
Also, we analyse the experimental data from RHIC and LHC with calculated nuclear modification factor. 
It is observed that the calculated $R_{AA}$ explains accurately for heavy flavor particles in all $p_T$ range, 
but it can only explain $R_{AA}$ for light flavor particles in intermediate to high-$p_T$ range with an exception for kaons, where the enhancement 
at high-$p_T$ can not be explained with the present model. 
The relaxation time, as found from data, is higher for heavy particles.

\section*{Acknowledgements}
ST acknowledges the financial support by DST INSPIRE program of Govt. of India. TB acknowledges the discussion with Dr. Santosh K. Das.

\begin{table*}

\centering


\caption { Centrality, ${\chi ^2}/{ndf}$ and different extracted parameters
after fitting Eq.~\ref{eq10} to the $R_{AA}$ data of different
particles for Pb+Pb collisions and Au+Au collisions at $\sqrt{\mathrm{s}_{NN}}$= 2.76 TeV
and $\sqrt{\mathrm{s}_{NN}}$= 200 GeV, respectively.}

\noindent\begin{tabular}{ |c|c|c|c|c|c| }

\hline
\multicolumn{6}{|c|}{\bf{Pb+Pb 2.76 TeV}}\\
\hline
Particle & Centrality(\%) & ${\chi ^2}/{ndf}$ & $t_f/\tau$ & $q$  & T (GeV) \\
\hline
$\pi ^+ + \pi ^-$        & 0-5   & 0.364461  & 2.07313 $\pm$ 0.061906  & 1.00151
$\pm$ 0.00149 & 0.17854 $\pm$ 0.00143  \\
$K^+ + K^-$              & 0-5   & 0.390686  & 2.24302 $\pm$ 0.098624  & 1.00406
$\pm$ 0.00125 & 0.16803 $\pm$ 0.00133  \\
$K_s^0$                  & 0-5   & 0.284477  & 1.88499 $\pm$ 0.100713  & 1.00410
$\pm$ 0.00373 & 0.17320 $\pm$ 0.00379  \\
$p+\bar p$    	         & 0-5   & 0.267087  & 1.72079 $\pm$ 0.163273  & 1.00378
$\pm$ 0.00079 & 0.15771 $\pm$ 0.00111  \\
$\Lambda + \bar \Lambda$ & 0-5   & 0.017809  & 1.85201 $\pm$ 0.649860  & 1.00600
$\pm$ 0.00369 & 0.15411 $\pm$ 0.00512  \\
$D^0$                    & 30-50 & 0.262131  & 0.84223 $\pm$ 0.099520  & 1.01915
$\pm$ 0.01202 & 0.13538 $\pm$ 0.01568  \\
$J/\psi$                 & 0-90  & 0.155083  & 1.06248 $\pm$ 0.157049  & 1.01252
$\pm$ 0.00998 & 0.14676 $\pm$ 0.01408  \\

\hline
\multicolumn{6}{|c|}{\bf{Au+Au 200 GeV}}\\
\hline
$\pi ^0$                & 0-5    & 0.24223   & 1.66530 $\pm$ 0.05601   & 1.0050
$\pm$ 0.00525 & 0.17483 $\pm$ 0.00550 \\
$D^0$                   & 0-10   & 2.27963   & 1.37833 $\pm$ 0.63285   & 1.0076
$\pm$ 0.00594 & 0.15273 $\pm$ 0.00688 \\
\hline

\end{tabular}

\label{table1}



\end{table*}


\end{document}